\pdfoutput=1
\documentclass[%
 reprint,
amsmath,amssymb,
aps,
prl
]{revtex4-1}

\usepackage{graphicx}
\usepackage{dcolumn}
\usepackage{bm}

\usepackage{amsmath,amssymb,amsbsy,amstext,amsthm,simplewick,amsfonts,graphicx}
\usepackage{mathrsfs}
\usepackage{graphicx}
\usepackage{wrapfig}
\usepackage{upgreek}
\usepackage{bm} 
\usepackage{framed}
\usepackage{bbm}
\usepackage{textcomp}
\usepackage{adjustbox}
\usepackage{makecell}
\usepackage{tcolorbox}
\usepackage{empheq}
\usepackage[normalem]{ulem}
\usepackage{enumitem}
\usepackage{braket} 
\usepackage{array}
\usepackage{dsfont}
\usepackage{ulem}
\usepackage{xcolor}
\usepackage{caption}
\usepackage{subcaption}
\usepackage{tocloft}
\usepackage{tikz}
\usepackage{xcolor}
\usepackage{leftindex}
\usetikzlibrary{patterns}

\usepackage{hyperref}
\hypersetup{colorlinks=true,
	breaklinks=true,
	pdfstartview=Fit,
	linkcolor=blue2,
	citecolor=red2,
	urlcolor=blue
}

\definecolor{red2}{RGB}{214, 39, 40}
\definecolor{green2}{RGB}{0,170,0}
\definecolor{blue2}{RGB}{0,100,200}
\definecolor{magenta2}{RGB}{191,64,191}
\definecolor{purple2}{RGB}{112,48,160}
\definecolor{orange2}{RGB}{255,192,0}

\graphicspath{{Fig/}}

\begin{document}

\preprint{APS/123-QED}

\title{Soft unification of exceptional effective field theories in de Sitter space}

\author{Zong-Zhe Du}
\email{zongzhe.du@nottingham.ac.uk}
\affiliation{
School of Physics and Astronomy,
			University of Nottingham, University Park, \\ Nottingham, NG7 2RD, UK}

\date{\today}

\vspace{20mm}

\begin{abstract}
\vspace{10mm}

We uncover a surprising universal soft behaviour (USB) of a de Sitter (dS) S-matrix for \textit{all} exceptional effective field theories (EFTs) in dS space, stemming from the recently proposed generalised energy conservation (GEC) condition. At leading order in the soft limit, the dS S-matrices for all exceptional EFTs in four spacetime dimensions exhibit no scaling with the soft momentum, i.e. $\lim_{k\rightarrow 0}\mathcal{A}(k)\sim\mathcal{O}(1)$, a feature we term USB. We specifically show that USB fixes the four-point amplitudes in Dirac-Born-Infeld (DBI) theory and Special Galileon, and six-point amplitude in $SU(N)$ Non-Linear Sigma Model (NLSM) given the minimal derivative count. We then conjecture USB fixes the interactions of all exceptional EFTs to all orders, thereby providing a unification criterion for them. Our result further underpins the idea that $\Delta\geq 4$ exceptional theories in dS are characterised by the spectrum and stability requirement alone, resonating with the low energy theory of gravity--General Relativity.
\end{abstract}

\maketitle

\newpage

\newpage

\section{Introduction}
The recently proposed Generalised Energy Conservation (GEC) for de Sitter (dS) scattering amplitude defined on an extended Poincare patch serves as a powerful tool to constrain theories living in dS space \cite{Donath:2024utn,bsyw-czt5}. The extended Poincare patch is defined by analytically continuing the conformal time $\eta$ from $(-\infty,0)$ to $(-\infty, + \infty)$ { which is mandated by the realisation of the full de Sitter isometry, an anti-de Sitter version of the argument can be found in \cite{Fujisawa:2014mha}}, therefore in and out states can be consistently prepared on the far past and far future boundary. Since there's no global time translation symmetry in (extended) dS space, energy conservation shouldn't be imposed as a prior condition. GEC thus refers to the vanishing requirement of the non-energy conserving part of the amplitude
\begin{align}
    \mathcal{A}_{k_T\neq 0} = 0,
\end{align}
where $k_T \equiv \sum_{a\in\text{in}} |\vec{k}|_{a} - \sum_{b\in\text{out}} |\vec{k}|_{b}$
 adhering to cosmologists' convention. Given the exceptional series in four-dimensional dS group representation theory \cite{Basile:2016aen,Sun:2021thf} where the conformal dimension $\Delta \geq 4$ is an integer, and the corresponding mass spectrum is given by $ m^2 = -\Delta(\Delta - 3)H^2$, it was shown in \cite{bsyw-czt5} that the scalar four-point dS amplitude can be uniquely fixed by imposing GEC \footnote{Unlike in flat space, the four-point dS amplitude cannot be uniquely fixed by the spectrum, derivative count, and dS invariance, since scale invariance allows different operators to contribute to the same kinematic dependence.}, which can be simply regarded as the requirement for the absence of instability within the theory \footnote{The non-energy conserving part of the amplitude only has support when energy in the future is greater (lesser) than energy in the past, which implies eternal energy creation(annihilation), leading to potential instability.}. It is therefore conjectured the full theory could be uniquely fixed by the GEC condition given a particular exceptional series mass spectrum. The theories that satisfy GEC are precisely the exceptional effective field theories (EFTs) stemmed from symmetries and brane world constructions \cite{Goon:2011qf,Bonifacio:2018zex,Bonifacio:2021mrf}.

Scalar theories that carry long range force in flat space, however, cannot be uniquely fixed by the spectrum, additional conditions such as soft limits \cite{Adler:1965ga,Cheung:2014dqa, Cheung:2015ota} and non-linear symmetries \cite{Hinterbichler:2015pqa,Roest:2019oiw} are entailed for the sake of pinning down the unique interactions. For instance, Dirac-Born-Infeld (DBI) is the unique $P(X)$ theory where $X = (\partial\phi)^2$ and $P(X)$ is a polynomial of $X$ that enjoys an enhanced Adler zero $\mathcal{A}(p)\sim\mathcal{O}(p^2)$, i.e. the scattering amplitude of DBI vanishes faster than naively expected upon sending one momentum soft. The enhanced Adler zeros are inextricably related to the non-linear symmetries \cite{Du:2024hol, Cheung:2016drk, Bittermann:2022nfh, Roest:2019oiw}. The soft scaling $\mathcal{A}(p)\sim\mathcal{O}(p^\sigma)$ provides a classification scheme for EFTs in flat space \cite{Cheung:2016drk}, where $\sigma = 2$ corresponds to DBI, $\sigma = 3$ corresponds to special Galileon, and no single scalar theoy exists beyond $\sigma = 3$. The same theories are also studied in the context of the wave function coefficients in flat space \cite{Bittermann:2022nfh} and in de Sitter space \cite{Armstrong:2022vgl}. The exceptional EFTs wherein the interactions are fixed up to a single coupling constant in dS space, thereby appear more fundamental than their flat space counterparts, since we need nothing more than the spectrum and stability requirement to fix the dS theories. 

In the present work, we bridge the two criteria used to classify EFTs in dS and flat space by examining the soft behaviour of dS amplitudes for the exceptional EFTs. Surprisingly, unlike in flat space where different theories exhibit different soft limits, we find that the exceptional EFTs in {four-dimensional} dS space enjoy a universal soft behaviour (USB), where the dS amplitudes exhibit no scaling with the soft momentum:
\begin{align}
    \lim_{k\rightarrow 0}\mathcal{A}(k)\sim\mathcal{O}(1). \label{USB}
\end{align}
This is the main result of the paper. USB can be derived from Ward-Takahashi identities with respect to known non-linear symmetries of the exceptional theories as is shown in the Supplementary Material \footnote{See Supplementary Material at [url] for a detailed derivation of Universal Soft Behaviour from Ward-Takahashi identity, which includes reference \cite{Hui:2018cag,Du:2024soq}.}, thus making it a non-perturbative property of the dS amplitudes. We also prove USB for four-point dS amplitudes by imposing GEC for $\Delta\geq 4$ theories.  Conversely, one can also use USB to uniquely bootstrap the four-point dS amplitudes. The non-energy conserving part of the dS amplitude $\mathcal{A}_{k_T\neq 0}$ violates USB for all $\Delta \geq 4$ exceptional EFTs, therefore imposing GEC by demanding the vanishing of $\mathcal{A}_{k_T \neq 0}$, is equivalent to imposing USB. In short, for $\Delta \geq 4$ exceptional EFTs, we have
\begin{align}
    \text{GEC} \Leftrightarrow \text{USB}.
\end{align}
Moreover, it goes without saying, USB and GEC are robust against any perturbative field redefinition since they're properties of the scattering amplitude. Practically, one needs to invoke highly non-trivial distribution identities involving Dirac-Delta and its derivatives to verify the field basis independence. The verification is done for all exceptional theories by showing amplitudes computed from diverse field basis converge to an identical expression for each theory.

A subtlety encountered in \cite{bsyw-czt5} is that for $\Delta = 3$ or massless scalar theories, GEC doesn't pick out the $SU(N)$ Non-Linear Sigma Model (henceforth referred to as NLSM) \cite{Cronin:1967jq, Weinberg:1966fm}, which describes the low energy dynamics of the Nambu-Goldstone mode corresponding to the chiral symmetry breaking $SU(N)\times SU(N)/SU(N)$. The on-shell avatar of flat space $SU(N)$ NLSM is extensively studied in \cite{Adler:1965ga, Paton:1969je, Ellis:1970nt,Susskind:1970gf, Kampf:2013vha, Bartsch:2024ofb, Kampf:2012fn, Rodina:2016jyz}. GEC \cite{bsyw-czt5} suggests that \textit{any} two derivative massless scalar theory appears to be healthy from the dS S-matrix point of view. The result is sensible, as the coset of $SU(N)$ NLSM is blind to the background space-time geometry it's living in. This further suggests that DBI and Special Galileon are more fundamental than $SU(N)$ NLSM in dS, where the former two theories can be uniquely determined by their spectrum and GEC (stability requirement) alone. USB, however, should fix the dS amplitude in $SU(N)$ NLSM as a consequence of the non-linearly realised symmetry.

We show that the six-point partial amplitude in a generic two-derivative flavour-ordered scalar theory is uniquely fixed by the four-point coupling constant upon imposing USB and that the resulting theory is precisely $SU(N)$ NLSM. The USB bootstrap thus complements the GEC conditions proposed in \cite{bsyw-czt5}, as we're able to include $\Delta = 3$ to our family of exceptional theories. 

\section{Notation}
We denote $k \equiv |\vec{k}|$ as the magnitude of 3-momentum, which we often refer to as the energy of the exceptional series scalars, $s_{ij} = -2 k_i k_j + 2\vec{k}_i \cdot\vec{k}_j$ as the Mandelstam variables, $H$ as the Hubble constant and $R = 1/H$ the Hubble radius.

\section{Universal Soft Behaviour from Generalised Energy Conservation}
 We start by reviewing the dS S-matrix construction proposed in \cite{Donath:2024utn} and GEC imposition for $\Delta\geq 4$ exceptional EFTs in dS \cite{bsyw-czt5}. The main idea is to analytically continue the conformal time $\eta$ defined on the Poincare patch of dS space from $(-\infty, 0)$ to $(-\infty, +\infty)$. The metric in the extended patch thus takes the following form
 \begin{align}
     ds^2 = \frac{-d\eta^2 + dx^2}{H^2\eta^2},\;\;\eta\in(-\infty, +\infty),
 \end{align}
where the $\eta = 0$ singularity requires special care as we will explain later in the section. The benefit of defining such extension is that we are granted with the privilege to prepare asymptotic free states in both the far past and far future, such that scattering experiments can be consistently performed. The dS S-matrix is computed by the Dyson series
 \begin{align}
     S_{\beta\alpha} = \braket{\beta|\mathcal{T}\{e^{-i \int \mathcal{H}_{\text{int}} d\eta} \}|\alpha} \equiv 1+ i(2\pi)^4 \mathcal{A} \delta^3(\Vec{k}_\beta - \Vec{k}_\alpha),
 \end{align}
where $\beta$ and $\alpha$ are out and in states in the interaction picture respectively, $\mathcal{H}_{\text{int}}$ is the interaction Hamiltonian in the interaction picture and $\mathcal{T}$ is the time ordering operator. The scalar mode expansion {in $D=4$} is given by
\begin{align}
        \phi(x) = \int \frac{d^3 k}{(2\pi)^3} f_k(\eta) e^{i\vec{k}\cdot\vec{x}} a_{\vec{k}} + c.c.,
\end{align}
where $a_{\vec{k}}$ is the annihilation operator, and the mode function $f_{k}(\eta)$ is given by
\begin{align}
    f_k(\eta) = \frac{\sqrt{\pi}H}{2} (-\eta)^{\frac{3}{2}} e^{i\frac{\pi}{2}(\nu - 1/2)} H_{\nu}^{(1)}(-k\eta),\label{ModeFunction}
\end{align}
where $\nu = \Delta - \frac{3}{2}$ and $H_{\nu}^{(1)}(x)$ is the Hankel function of the first kind. For exceptional series scalars with integer $\Delta\geq 3$, the mode function has no branch cut. Therefore for such theories, the interaction Hamiltonian in the Dyson expansion only contains Laurent polynomials of $\eta$, leading to the following dS amplitude ansatz
\begin{align}
    \mathcal{A} = \mathcal{A}^{(\pm)}_{k_T \neq 0} + \sum_{m} \mathcal{A}^{(m)}\delta^{(m)}(k_T),\label{Ansatz}
\end{align}
where $\delta^{(m)}(k_T)$ denotes the $m$ th derivative of $\delta(k_T)$. The non-energy conserving part of the amplitude $\mathcal{A}_{k_T\neq 0}^{(\pm)}$ emanates from the part of the integrand that diverges at $\eta = 0$. The upper index $(\pm)$ denotes two choices of time integration contour deformation at $\eta = 0$, one goes above the pole and one goes below. The integration contour is explicitly shown in Fig 1 and Fig 2 in \cite{bsyw-czt5}. If the contour goes above the pole, then the time integral only has support when $k_T \geq 0$, namely eternal energy annihilation; if the contour goes below the pole, then the time integral only has support when $k_T \leq 0$, namely eternal energy creation. In either case, the non-energy conserving part $\mathcal{A}^{(\pm)}_{k_T\neq 0}$ could lead to potential instability, therefore the GEC condition of \cite{bsyw-czt5} is given by demanding the absence of potential instability
 \begin{align}
     \mathcal{A}^{(\pm)}_{k_T\neq 0} = 0. \label{GEC}
 \end{align}
As we shall see, the GEC condition strongly constrains the coupling constants of the theories, leading to known exceptional EFTs and evidence for new theories. In the following examples, we study the consequence of GEC imposition on soft behaviours of four-point amplitudes in $\Delta\geq 4$ exceptional theories, while postponing the discussion of $\Delta = 3$ case to the next chapter.
\paragraph{$\bullet\Delta = 4$}
The generic Lagrangian for $\Delta = 4$ theory up to quartic interaction and four derivatives is given by
\begin{align} \label{FourDerivativeLagrangian}
    \frac{\mathcal{L}^{\Delta = 4}}{\sqrt{-g}} &= -\frac{1}{2} (\partial\phi)^2 + 2 H^2 \phi^2 + d^{(4)}_0 H^4 \phi^4 + d^{(4)}_4 (\partial \phi)^4 \ldots
\end{align}
The non-energy conserving part of the amplitude reads \cite{bsyw-czt5}
\begin{align}
    &\mathcal{A}^{(\pm)}_4|_{k_T\neq 0} = -(d_0^{(4)} + 8 d_4^{(4)}) H^4 i\frac{36}{35}  \theta(\pm k_T) \nonumber\\
    &\frac{\biggl( 3 ( k_1^7 + k_2^7 +k_3^7 +k_4^7) - 7[k_1^2 k_2^2 (k_1^3 + k_2^3) + 5\text{perms} ] \biggr)}{k_1^2 k_2^2 k_3^2 k_4^2}\nonumber\\
    &= (d_0^{(4)} + 8 d_4^{(4)}) K_{1, k_T\neq 0}^{\Delta = 4} \times (k_1^{-2}),\label{Delta=4NEC}
\end{align}
where $K_{a,k_T\neq 0}^{\Delta = 4}$ denotes the kinematic dependence regular in $k_1$, i.e.
\begin{align}
    \lim_{k_1\rightarrow 0}K_{a,k_T\neq 0}^{\Delta = 4}\sim\mathcal{O}(1),
\end{align}
it exhibits no scaling with the soft momentum at leading order in $k_1$. The subscript $a$ simply serves as a label. GEC requires a vanishing $\mathcal{A}^{(\pm)}_{k_T\neq 0}$, leading to the tuning $d^{(4)}_0 = -8 d^{(4)}_4$, which corresponds to the DBI action \cite{Bonifacio:2021mrf, bsyw-czt5} in dS space. The DBI four-point amplitude has the following soft behaviour
\begin{align}
    \mathcal{A}_4^{\text{DBI}} &= K_0^{\Delta = 4} \delta(k_T)+K_1^{\Delta = 4}\delta^{(1)}(k_T) \nonumber\\
    &+ K_2^{\Delta = 4}\delta^{(2)}(k_T) + K_3^{\Delta = 4}\times (k_1) \delta^{(3)}(k_T)\nonumber\\
    &+K_4^{\Delta = 4} \times (k_1^2) \delta^{(4)}(k_T).
\end{align}
The $K_a^{\Delta = 4}$ here again denotes kinematic dependence that is regular and starts at $\mathcal{O}(1)$ in the soft momentum expansion. We therefore see the GEC condition implies the soft behaviour
\begin{align}
    \lim_{k_1\rightarrow 0}\mathcal{A}(k_1)\sim\mathcal{O}(1).
\end{align}
Conversely, since the non-energy conserving part of the amplitude scales with the soft momentum as
\begin{align}
    \lim_{k_1\rightarrow 0}\mathcal{A}^{(\pm)}_{k_T}\sim \mathcal{O}(k_1^{-2}),
\end{align}
imposing the soft behaviour implies GEC as \eqref{Delta=4NEC} violates USB. Therefore we've shown at the level of four-point interactions, GEC is equivalent to USB.
\paragraph{$\bullet\Delta = 5$}
The generic Lagrangian for $\Delta = 5$ theory up to quartic interaction and six derivatives is given by
\begin{align}
    \frac{\mathcal{L}^{\Delta = 5}}{\sqrt{-g}} = &-\frac{1}{2}(\partial\phi)^2 + 5 H^2 \phi^2 + d^{(5)}_{0}H^6 \phi^4 \nonumber \\
    &+ d_{4}^{(5)} H^2 (\partial\phi)^4 + d_{6}^{(5)} (\partial\phi)^2 (\nabla_\mu\nabla_\nu\phi)^2 + \ldots
\end{align}
The non-energy conserving part of the amplitude has the following soft behaviour \cite{bsyw-czt5},
\begin{align}
    \mathcal{A}_{k_T\neq 0}^{(\pm)} &= \biggl((d_0^{(5)} + \frac{125}{6}d_4^{(5)})K^{\Delta = 5}_{1,k_T\neq 0} \nonumber\\
    &+ (d_0^{(5)} - \frac{500}{3}d_6^{(5)})K^{\Delta = 5}_{2,k_T\neq 0}\biggr)\times (k_1^{-3}),
\end{align}
Again imposing GEC is equivalent to imposing USB \eqref{USB} and vice versa. Both criteria lead to the theory of Special Galileon in dS space \cite{Bonifacio:2021mrf, Armstrong:2022vgl}.
\paragraph{$\bullet\Delta \geq 6$}
Despite not knowing the full spectrum for these new theories found in \cite{bsyw-czt5}, we do have full control of the scalar four-point self interaction. We find the non-energy conserving part of the amplitude diverges in the soft limit for all exceptional series scalars ($\Delta\geq 3$) with generic interactions as
\begin{align}
    \lim_{k\rightarrow 0}\mathcal{A}^{(\pm)}_{k_T\neq 0}\sim \mathcal{O}(k^{-\Delta+2}),\label{SoftScaling}
\end{align}
which can be schematically understood by the soft expansion of the mode function \eqref{ModeFunction} 
\begin{align}
    \lim_{k\rightarrow 0} f_k(\eta)\propto \frac{1}{\sqrt{k} } k^{-\Delta + 2},
\end{align}
where the normalisation factor $\frac{1}{\sqrt{k}}$ is cancelled by the $\sqrt{k}$ factor from the action of creation(annihilation) operator on the vacuum state when computing the dS amplitudes, i.e.
\begin{align}
    \sqrt{2k} a^\dag_{\vec{k}}\ket{0} =  \ket{\vec{k}}.
\end{align}
Therefore both GEC and USB are violated if the interactions are kept generic (without tuning). The imposition of GEC and USB thus lead to the same unique coupling tuning since they both require a vanishing $\mathcal{A}^{(\pm)}_{k_T\neq 0}$. Remarkably, the unique tuning consistently results in USB \eqref{USB} for the energy-conserving part of the amplitudes, for which we present a symmetry based proof in the supplementary material.

\section{Non-Linear Sigma Model at Six-Point}

The Lagrangian of the $SU(N)$ NLSM can be written as \cite{Cronin:1967jq}
\begin{align}
    \frac{\mathcal{L}}{\sqrt{-g}} = -\frac{f_\pi^2}{4}\text{Tr}[\partial_\mu U \partial^\mu U^\dag],
\end{align}
where $f_\pi$ is the pion decay constant, and the unitary matrix $U$ is parametrised by
\begin{align}
    U = \text{exp}\biggl(i \sqrt{2}\frac{\pi_a}{f_\pi} T_a \biggr),
\end{align}
with $\pi_a$ denoting the pion fields and $T_a$ the generators of the coset $SU(N)\times SU(N)/SU(N)$. The equivalence between GEC and USB fails to hold for $SU(N)$ NLSM ($\Delta = 3$). Since the massless mode function
\begin{align}
    f_k(\eta) = \frac{H e^{-ik\eta}}{\sqrt{2k^3}}(1 + ik\eta),
\end{align}
lacks a term linear in $\eta$ \cite{MLT} in the expansion around $\eta = 0$, the time integrand in any derivatively coupled massless scalar theory does not contain a $\frac{1}{\eta}$ factor, and the non–energy-conserving part of the amplitude, $\mathcal{A}_{k_T\neq 0}^{(\pm)}$, which arises from the residue of the time integrand, always vanishes. A detailed analysis of the vanishing residue of the time integrand is presented in \cite{bsyw-czt5}. This suggests $SU(N)$ NLSM cannot be uniquely fixed by spectrum and GEC. However, owing to the presence of the non-linear symmetry, $SU(N)$ NLSM exhibits the same USB as other exceptional theories, we should be able to fix the theory by imposing the USB condition. In \cite{bsyw-czt5}, it was shown that for massless scalars, the presence of any potential term---regardless of additional interactions---violates GEC and hence USB according to \eqref{SoftScaling} even for $\Delta = 3$, therefore we only consider scalar theories with two derivatives in this section. 

 The tree-level amplitude for a generic massless two-derivative scalar theory equipped with a flavour-ordered group decomposition structure\cite{Paton:1969je,Susskind:1970gf,Ellis:1970nt,Kampf:2012fn, Kampf:2013vha} is given by
\begin{align}
    \mathcal{A}_n = \sum_{\text{non-cyclic perms}} \text{Tr}(T^{a_1}\ldots T^{a_n}) \mathcal{M}(\vec{k}_{a_1},\ldots,\vec{k}_{a_n}),
\end{align}
where $\mathcal{M}(\vec{k}_{a_1},\ldots,\vec{k}_{a_n})$ is referred to as the \textit{partial amplitude}. The decomposition is unmodified in dS space since the symmetry is strictly internal. The four-point partial amplitude are computed by summing over $s_{12}$ and $s_{14}$ invariant contributions according to the Feynman rules proposed in \cite{Kampf:2013vha} while putting the momenta on-shell, i.e.
\begin{align}
    \mathcal{M}_4 = \mathcal{M}^{(0)}_4 \delta(k_T) +  \mathcal{M}^{(1)}_4 \delta^{(1)}(k_T) + \mathcal{M}^{(2)}_4 \delta^{(2)}(k_T),
\end{align}
where the amplitude coefficients are given by
\begin{align}
    \mathcal{M}^{(0)}_4 &= g_{4}\frac{2}{k_1 k_2 k_3 (k_1 + k_2 + k_3)^2}\nonumber\\
    &\biggl(k_1^2 (k_3^2 + \vec{k}_1\cdot\vec{k}_3)(k_1 + 2 k_2 + k_3) \nonumber\\
    &+ \vec{k}_1\cdot\vec{k}_3[(2k_1 + k_2 + k_3)(k_2^2 + k_2 k_3 + k_3^2) {+ k_1^2 k_3]}\biggr) ,\label{4ptamp0}
    \end{align}
    \begin{align}
        \mathcal{M}^{(1)}_4 &= g_{4}\frac{2}{k_1 k_2 k_3 (k_1 + k_2 + k_3)}\nonumber\\
    &\biggl((k_1 + k_3) [k_1^2 k_3^2 - \vec{k}_1\cdot\vec{k}_3 (k_1 + k_2)(k_2 + k_3)] \biggr),\label{4ptamp1}
    \end{align}
    \begin{align}
    \mathcal{M}^{(2)}_4 &= g_{4} s_{13}. \label{4ptamp2}
\end{align}
The four-point coupling $g_4$ is related to the pion decay rate by $g_4 = \frac{H^2}{2f_\pi^2}$. It can be easily checked that USB is trivially satisfied here, analogous to the case where the four-point amplitude in flat space $SU(N)$ NLSM trivially satisfies Adler zero by momentum conservation and derivative count.

The six-point amplitude can be split into the contact contribution and exchange contribution
\begin{align}
    \mathcal{M}_6 = \mathcal{M}_6^{\text{contact}} + \mathcal{M}_6^{\text{exchange}}.
\end{align}
The contact term can also be directly computed via the Feynman rules and putting the momenta on-shell. The full expression of the amplitude is unwieldy, here we present only the leading order soft behaviour, which is the main focus of our analysis
\begin{align}
    &\;\;\mathcal{M}_6^{(0)\text{contact}}\nonumber\\
    &= g_{6}\frac{2}{ k_2 k_3 k_4 k_5} \biggl[ -k_2^3 - k_3^3 + ( -\vec{k}_2 \cdot \vec{k}_3 + \vec{k}_2 \cdot \vec{k}_5 \nonumber\\
    &- \vec{k}_3 \cdot \vec{k}_4 + \vec{k}_3 \cdot \vec{k}_5 + k_2^2 + k_3^2 ) \biggl(k_5 + \frac{k_2^2 + k_3^2 + k_4^2}{k_1 + k_2 + k_3 + k_4 + k_5}\biggr)\nonumber\\ 
    &+ \frac{k_2^4 + k_3^4 - 3 k_3^2 k_4^2 - 4 k_2^2 k_3^2 -k_2^2 k_4^2}{k_1 + k_2 + k_3 + k_4 + k_5}\biggr] \times (k_1^{-1}) + \mathcal{O}(1)
    \end{align}
    \begin{align}
    &\;\;\mathcal{M}_6^{(1)\text{contact}}\nonumber\\
    &= g_{6}\frac{2}{ k_2 k_3 k_4 k_5} \biggl[ k_2 k_3 (k_2 + k_3)^2 + (k_2^3 + k_3^3) k_4 \nonumber\\
    &+ (k_2^2 + k_3^2) (k_4^2 + k_5^2)\nonumber\\
    &-\biggl( (k_2 + k_3 + k_4)(-\vec{k}_2\cdot\vec{k}_3 + \vec{k}_2 \cdot\vec{k}_5 - \vec{k}_3 \cdot\vec{k}_4 + \vec{k}_3 \cdot\vec{k}_5) \nonumber\\
    &+ (k_2 + k_3) k_2 k_3 + (k_2^2 + k_3^2) k_4 \biggr) k_5 \biggr]\times (k_1^{-1}) + \mathcal{O}(1),
    \end{align}
    \begin{align}
    &\;\;\mathcal{M}_6^{(2)\text{contact}}\nonumber\\
    &= g_{6}\frac{2}{ k_2 k_3 k_4 k_5}\biggl[(-\vec{k}_2\cdot\vec{k}_3 + \vec{k}_2 \cdot\vec{k}_5 - \vec{k}_3 \cdot\vec{k}_4 + \vec{k}_3 \cdot\vec{k}_5)(k_2 k_3 k_4 \nonumber\\
    &+ k_2 k_3 k_5 + k_2 k_4 k_5 + k_3 k_4 k_5) + k_3^2  k_4^2 k_5\nonumber\\
    &+k_2 k_3^2 \left(k_2+k_4\right) k_4+k_2 k_3 \left(k_2+k_3\right) k_4 k_5\nonumber\\
    &-\left(k_2 k_3 \left(k_2+k_3\right)+\left(k_2^2+k_3^2\right) k_4\right) k_5^2+k_2^2 k_3^2 k_5 \biggr]\times (k_1^{-1}) \nonumber\\
    &+ \mathcal{O}(1),
    \end{align}
    \begin{align}
    &\;\;\mathcal{M}_6^{(3)\text{contact}}\nonumber\\
    &= - g_{6} (-s_{23} + s_{25} - s_{34} + s_{35}) \times (k_1^{-1}) + \mathcal{O}(1),
    \end{align}
\begin{align}
    \mathcal{M}_6^{(4)\text{contact}} = g_6 (-s_{23} + s_{25} - s_{34} + s_{35}) + \mathcal{O}(k_1),
\end{align}
where $g_6$ is the six-point coupling constant, the superscript $(p)$ in $\mathcal{M}^{(p)}_n$ denotes the amplitude coefficient index associated with the number of derivatives on $\delta(k_T)$, and the subscript $n$ denotes the number of external legs. The exchange contribution is computed by an exchange energy integral
\begin{align}
    \mathcal{M}_6^{\text{exchange}} =& \int_{-\infty}^{+\infty} \frac{d\omega}{\omega^2 - (\vec{k}_1 + \vec{k}_2 + \vec{k}_3)^2}V_{123}(\omega)V_{456}(-\omega) \nonumber\\
    &+ \text{cyclic},\label{6ptExchange}
\end{align}
where the propagator we employed here is the usual Feynman propagator proposed in \cite{Donath:2024utn} up to a normalisation factor. The four-point vertex $V_{abc}$ can be decomposed into an energy delta function and its derivatives thereof according to the general ansatz \eqref{Ansatz}
\begin{align}
    V_{abc}(\omega) = \sum_{m=0}^{2}V_{abc}^{(m)}\delta^{(m)}(\omega + k_a + k_b + k_c).
\end{align}
The vertex coefficients are given by
\begin{align}
    &\;\;V_{abc}^{(0)} \nonumber\\
    &= g_{4}\biggl[\frac{2 k_b^2}{k_a k_c} -\frac{2}{k_a k_b k_c (k_a + k_ b + k_c)^2} (\vec{k}_a \cdot \vec{k}_b + \vec{k}_b \cdot \vec{k}_c + k_b^2 )\nonumber\\
    &\biggl(k_a^3 + 2 k_a^2 (k_b + k_c) + (2 k_a + k_b + k_c)(k_b^2 + k_b k_c + k_c^2)\biggr)\biggr],
\end{align}
\begin{align}
    &\;\;V_{abc}^{(1)} \nonumber\\
    &= g_{4}\frac{2}{k_a k_b k_c (k_a + k_b + k_c)}\biggl((\vec{k}_a \cdot \vec{k}_b + \vec{k}_b \cdot \vec{k}_c)(k_a + k_b)\nonumber\\
    &\;\;(k_a + k_c)(k_b + k_c) - k_b^2 [k_a^2 (k_a + k_b) + k_c^2 (k_b + k_c)] \biggr),
\end{align}
\begin{align}
    &\;\;V_{abc}^{(2)} = -g_{4}(s_{ab} + s_{bc}).
\end{align}
Notice the four-point amplitude \eqref{4ptamp0}-\eqref{4ptamp2} can be obtained by evaluating the four-point vertex $V_{123}(\omega)$ in the on-shell limit
\begin{align}
    \omega^2 = (\vec{k}_1 + \vec{k}_2 + \vec{k}_3)^2,
\end{align} 
as expected from factorisation. Upon integration by part, one can repackage \eqref{6ptExchange} into the same form as the contact term abiding to the general ansatz \eqref{Ansatz}. 
Both $\mathcal{M}_6^{\text{contact}}$ and $\mathcal{M}_6^{\text{exchange}}$ diverge in the soft limit and scale as $\mathcal{O}(k^{-1})$. Therefore a collaboration between the two is required for USB to hold, analogous to the mechanism ensuring (enhanced) Adler zeros in flat space $SU(N)$ NLSM, DBI and special Galileon. For simplicity, let's evaluate the amplitude coefficient associated with $\delta^{(3)}(k_T)$, the exchange contribution is obtained via integration by part
\begin{align}
&\mathcal{M}_6^{(3)\text{exchange}} = -g_4^2 \frac{1}{(k_1 + k_2 + k_3)^2 - (\vec{k}_1 + \vec{k}_2 + \vec{k}_3)^2} \nonumber\\
&\biggl(V_{123}^{(1)}V_{456}^{(2)} + V_{456}^{(1)} V_{123}^{(2)} + \frac{4(k_1 + k_2 + k_3)V_{123}^{(2)}V_{456}^{(2)}}{(k_1 + k_2 + k_3)^2 - (\vec{k}_1 + \vec{k}_2 + \vec{k}_3)^2}\biggr)\nonumber\\
&+ \text{cyclic}.
\end{align}
The overall contribution from both contact and exchange terms is then
\begin{align}
    \mathcal{M}_6^{(3)} &= \mathcal{M}_6^{(3)\text{contact}} + \mathcal{M}_6^{(3)\text{exchange}} \nonumber\\
    &=(-g_6 + g_4^2)(-s_{23} + s_{25} - s_{34} + s_{35}) \times (k_1^{-1}) \nonumber\\
    &+ \mathcal{O}(1).
\end{align}
 USB thus requires the following coupling tuning
\begin{align}
    g_6 = g_4^2,
\end{align}
which precisely corresponds to $SU(N)$ NLSM \cite{Cronin:1967jq, Weinberg:1966fm, Kampf:2013vha}. One can show imposing USB on $\mathcal{M}^{(0)}_{6}, \mathcal{M}^{(1)}_{6} \text{and}\; \mathcal{M}^{(2)}_{6}$ leads to the same coupling tuning in all cases. Therefore, we conclude the six-point dS amplitude in $SU(N)$ NLSM is completely fixed by imposing USB.

\section{Discussion}
In this work we have studied the soft limits of a dS S-matrix for \textit{all} exceptional EFTs in four spacetime dimensions and found they exhibit a universal soft behaviour \eqref{USB}, wherein the soft amplitude is regular in the soft momentum expansion. Specifically, we showed that USB is satisfied for all exceptional EFTs discovered in \cite{bsyw-czt5} at the level of four-point amplitudes, and imposing USB is equivalent to imposing GEC. We therefore conjecture the equivalence between USB and GEC holds to all orders for all $\Delta\geq 4$ exceptional theories. In addition, we showed that USB uniquely fixes the six-point partial amplitude in terms of four-point coefficients for a generic flavour-ordered two-derivative scalar theory, where the resulting theory precisely corresponds to $SU(N)$ NLSM. This leads us to conjecture that USB entirely fixes tree-level dS amplitudes for $SU(N)$ NLSM by recursion relations, along the same lines as the flat space analysis \cite{Susskind:1970gf,Ellis:1970nt,Kampf:2012fn,Cheung:2015ota, Rodina:2016jyz}. The proof of the conjecture requires our detailed knowledge of the analytical structure of the dS amplitude and the soft behaviour for each amplitude coefficient which are absent in our current analysis. To test the conjecture, a necessary step is to compute the six-point amplitudes for all exceptional theories. The technical obstruction is that except for NLSM, any other exceptional theories have explicit spacetime dimension dependence. Therefore we expect that to see the cancellation among different diagrams, it's necessary to invoke dimension dependent identities for the external momenta which are highly non-linear. Numerical implementation in accordance with \cite{Cheung:2014dqa} seems to be unavoidable. We will leave the higher-point test to future works.

Despite the universality across all exceptional theories, it's worth emphasising that the soft behaviour \eqref{USB} is not a necessary criterion for a theory to be healthy. Instead, it should be regarded as a distinctive feature of the exceptional theories. GEC, on the other hand, is a fundamental requirement for a healthy theory, as its violation could lead to potential instabilities. Theories that can be uniquely fixed by GEC, such as DBI and Special Galileon, are considered more fundamental than those requiring additional conditions such as USB, as in the case of $SU(N)$ NLSM, for their uniqueness. The former can be placed on the same footing as the low energy theory for gravity--General Relativity, where the theory is uniquely fixed by the spectrum and stability requirements alone \cite{Deser:1969wk}. It is nonetheless interesting to study theories that cannot be uniquely fixed by GEC but can be by USB, such as $SU(N)$ NLSM.

\section{Acknowledgement}
The author would like to thank Laurentiu Rodina and David Stefanyszyn for useful discussions. Z.D. is supported by Nottingham CSC [file No. 202206340007].

For the purpose of open access, the authors have applied a CC BY public copyright licence to any Author
Accepted Manuscript version arising.

\paragraph{Data access statement} No new data was
created or analysed during this study. 

\section{Appendix: Universal Soft Behaviour from Symmetries}

Here we derive USB of the dS amplitudes for exceptional theories with non-Abelian shift symmetries \cite{Bonifacio:2018zex, Bonifacio:2021mrf}, while focusing exclusively in $D=4$.

\subsection{$SU(N)$ NLSM}
The starting point is the charge conservation formula
\begin{align}
    \braket{\beta|Q S - S Q|\alpha} = 0, \label{ChargeConservation}
\end{align}
 where $Q$ is the conserved charge of the symmetry under consideration, $S$ is the S-matrix and $\beta,\alpha$ are out and in state in the interaction picture respectively. Consider the symmetry for $SU(N)$ NLSM \cite{Cronin:1967jq, Bittermann:2022nfh} in de Sitter space,
 \begin{align}
     \delta\phi = c + \mathcal{O}(\phi^2).
 \end{align}
 The conserved charge evaluated at asymptotic past(future) is given by three terms
\begin{align}
    Q^{\text{NLSM}} &=\int d^3 x J^0 = \int d^3 x R\frac{\partial_{\eta}\phi(x)}{\eta^2}+\mathcal{O}(\phi^3)
    \nonumber\\
    &= \int d^3 x \frac{1}{\eta}\int \frac{d^3 k}{(2\pi)^3} \sqrt{\frac{k}{2}} e^{i\vec{k}\cdot\vec{x}}  [a_{\vec{k}} e^{-ik\eta} + a_{-\vec{k}}^\dag e^{ik\eta}]\nonumber\\
    &+\mathcal{O}(\phi^3) \nonumber\\
    &= \lim_{\vec{k}\rightarrow 0} \sqrt{\frac{k}{2}}\frac{1}{\eta}[ a_{\vec{k}} e^{-ik\eta} + a_{-\vec{k}}^\dag e^{ik\eta}] +\mathcal{O}(\phi^3).
\end{align}
The $\mathcal{O}(\phi^3)$ doesn't contribute to the conserved charge since
\begin{align}
    &\int d^3 x \phi^3(x) = \int d^3 x \biggl(\int \frac{d^3 k_1}{(2\pi)^3} f_{k_1}(\eta) e^{i\vec{k}_1\cdot\vec{x}} a_{\vec{k}_1} + c.c.\biggr)^3 \nonumber\\
    &=\int_{k_1,k_2} e^{-i (k_1 + k_2 - k_{12})\eta}(1+ i k_1 \eta)(1 + i k_2\eta) (1- i k_{12}\eta)\nonumber\\
    &\;\;\;a_{\vec{k}_1} a_{\vec{k}_2} a_{-\vec{k}_1 - \vec{k}_2}^\dag + \ldots\stackrel{\eta\rightarrow \pm \infty,\; k_1\cdot k_2\neq 0}{\sim} 0.
\end{align}
The integrand highly oscillates in the far past (future) away from co-linear limit, leading to the vanishing of the integral over momentum space by $i\varepsilon$ prescription. A more detailed discussion can be found in \cite{Hui:2018cag}. Therefore, inserting the above expression of the Noether's charge back to the conservation equation \eqref{ChargeConservation} yields 
\begin{align}
   \lim_{\vec{k}\rightarrow 0,\eta\rightarrow -\infty,k\eta<< 1} &\frac{\braket{\beta,\vec{k}|S|\alpha}}{\eta_1} - \frac{\braket{\beta|S|-\vec{k},\alpha}}{\eta_2}  = 0,
\end{align}
where $\eta_1$ and $\eta_2$ are the far past(future) boundary, which do not have to be the same. Notice  that the limits we're taking here follow from \cite{Hui:2018cag} as well. The soft theorem we get is then
\begin{align}
  \lim_{k\rightarrow 0, k\eta<<1} \frac{k\braket{\beta,\vec{k}|\alpha}}{k\eta} = 0 \Rightarrow  \lim_{\vec{k}\rightarrow 0} k\braket{\beta,\vec{k}|S|\alpha}  &= 0,
\end{align}
where the two constraints from $\eta_1$ and $\eta_2$ converge into one by crossing symmetry. The solution to the soft theorem is given by
\begin{align}
   \lim_{k\rightarrow 0} \mathcal{A}^{\text{NLSM}}(k) \sim \mathcal{O}(1),
\end{align}
namely the soft amplitude does not scale with the soft momentum.

\subsection{DBI}
The DBI symmetry transformation in the ambient space reads \cite{Bonifacio:2021mrf}
\begin{align}
    \delta\Phi = S_A (X^A - \alpha \Phi\partial^{A}\Phi)
\end{align}
where $A= 0,1,\ldots,D$ and $\alpha$ is the deformation parameter. The ambient space coordinates are parametrised on the Poincare patch of $\text{dS}_4$ by
\begin{align}
    X^0 &= \frac{R^2 + x^2 -\eta^2}{2\eta},\;X^i = R\frac{x^i}{\eta},\;X^{D} = \frac{R^2 - x^2 + \eta^2}{2\eta};\\
    \eta &= \frac{R^2}{X^0 + X^D},\; x^i = R\frac{X^i}{X^0 + X^D}
\end{align}
The DBI symmetry variation in terms of the embedded space coordinates is then
\begin{align}
    \delta\phi =& S_i \biggl(\frac{R x^i}{\eta}\biggr) + \frac{S_0 + S_D}{2} \biggl(\frac{R^2}{\eta}\biggl)+ \frac{S_0 - S_D}{2} \biggr(\frac{x^2 -\eta^2}{\eta}\biggl)\nonumber\\
    &+ \mathcal{O}(\phi^2).
\end{align}
Therefore, there are $D$ non-linear symmetries in total. The DBI charge in $D=4$ is then given by
\begin{align}
    Q^{\text{DBI}} =& \int d^3 x J^0 = \int d^3 x \;\;S_+ R^4 \biggl(\frac{1}{\eta^3}\partial_\eta \phi + \frac{1}{\eta^4}\phi \biggr) \nonumber\\
    &+  S_i R^3 \biggl(\frac{x^i}{\eta^3}\partial_\eta \phi + \frac{x^i}{\eta^4}\phi\biggr) \nonumber\\
    & + S_- R^2 \biggl(\frac{x^2 - \eta^2}{\eta^3}\partial_\eta\phi + \frac{x^2 + \eta^2}{\eta^4}\phi \biggr) + \mathcal{O}(\phi^3),
\end{align}
where $S^+\equiv\frac{S_0 + S_D}{2}$ and $S^-\equiv\frac{S_0 - S_D}{2}$ are the 'light cone' symmetry parameters. The charge in terms of annihilation and creation operator is then
\begin{align}
    Q^{\text{DBI}} &= \lim_{k\rightarrow 0} S_+ R^3 \biggl[\frac{i}{\sqrt{k}\eta^3}(a_{\vec{k}} e^{-ik\eta} - a_{-\vec{k}}^\dag e^{ik\eta}) \nonumber\\
    &- \frac{\sqrt{k}}{\eta^2}(a_{\vec{k}}e^{-ik\eta} + a_{-\vec{k}}^\dag e^{ik\eta} ) \biggr] \\
    &+ i S_i R^2 \partial_{k^i}  \biggl[\frac{i}{\sqrt{k}\eta^3}(a_{\vec{k}} e^{-ik\eta} - a_{-\vec{k}}^\dag e^{ik\eta}) \nonumber\\
    &- \frac{\sqrt{k}}{\eta^2}(a_{\vec{k}}e^{-ik\eta} + a_{-\vec{k}}^\dag e^{ik\eta} ) \biggr]+\ldots
\end{align}
Here we omitted the the $S_-$ charge as it is degenerate with dS boost, $S^+$ and $S_i$. The soft theorems for $S_+$ and $S_i$ charges are then
\begin{align}
S_+\lim_{\vec{k}\rightarrow 0} k^2\braket{\beta,\vec{k}|S|\alpha} &= 0,\\
S_i\lim_{\vec{k}\rightarrow 0} \partial_{k_i} (k^2\braket{\beta,\vec{k}|S|\alpha}) &= 0.
\end{align}
The two set of soft theorems combine into a single solution for DBI amplitudes
\begin{align}
   \lim_{k\rightarrow 0} \mathcal{A}^{\text{DBI}} (k)\sim \mathcal{O}(1).
\end{align}

\subsection{Special Galileon}
For Special Galileon, the leading order symmetry transformation is
\begin{align}
    \delta\phi = S_{AB} X^A X^B + \mathcal{O}(\phi^2).
\end{align}
Followed by a similar derivation from DBI, the soft theorems are
\begin{align}
    S_{++} \lim_{k\rightarrow0} k^3 \braket{\beta,\vec{k}|S|\alpha} &= 0,\\
    S_{+i} \lim_{k\rightarrow0} \partial_{k^i} (k^3 \braket{\beta,\vec{k}|S|\alpha}) &= 0,\\
    S_{ij} \lim_{k\rightarrow 0} \partial_{k_i}\partial_{k_j}(k^3 \braket{\beta,\vec{k}|S|\alpha}) &= 0.
\end{align}
We therefore conclude
\begin{align}
    \mathcal{A}^{\text{sGal}} \sim \mathcal{O}(k^{0}).
\end{align}

\subsection{$\Delta\geq 6$}
The above analysis goes beyond known examples, here we postulates that the symmetry transformation takes the schematic form \cite{Bonifacio:2018zex}
\begin{align}
    \delta\phi = S_{A_1\ldots A_n}X^{A_1}\ldots X^{A_n} + \mathcal{O}(\text{field}^2),
\end{align}
where the non-Abelian term could comprise different spectrum content, but they necessarily start at quadratic order in fields. This leads to the absence of quadratic in fields term within the conserved charge. Notice the above statement is usually independent of any field basis choice \cite{Du:2024soq}. The schematic soft theorems following the previous constructions are then
\begin{align}
    S_{+\ldots +}\lim_{k\rightarrow 0} k^n \braket{\beta,\vec{k}|S|\alpha} &= 0,\\
    S_{+\ldots i}\lim_{k\rightarrow 0} \partial_{k_i} (k^n \braket{\beta,\vec{k}|S|\alpha}) &= 0,\\
    &\ldots\nonumber\\
    S_{i_1\ldots i_n}\lim_{k\rightarrow 0}\partial_{k_{i_1}}\ldots\partial_{k_{i_n}} (k^n \braket{\beta,\vec{k}|S|\alpha}) & = 0.
\end{align}
The solution to the soft theorems converges into
\begin{align}
    \lim_{k\rightarrow 0}\mathcal{A}(k) \sim \mathcal{O}(1).
\end{align}
Remarkably, even though the form of the soft theorems are different for each symmetry, the resulting soft behaviour for all exceptional EFTs in four spacetime dimensions is identical{\footnote{USB in each spacetime dimension should be different, the proof of which we defer to future work.}}, which drastically differs from their flat space counterparts.

\bibliographystyle{apsrev4-1}
\bibliography{RefDBI}

\end{document}